\documentclass[conference]{IEEEtran}
\IEEEoverridecommandlockouts
\usepackage{cite}
\usepackage{algpseudocode}
\usepackage{algorithm}
\usepackage{authblk}
\usepackage{graphicx}
\usepackage{textcomp}
\usepackage{xcolor}
\usepackage{subcaption}
\usepackage{amsmath,amssymb}

\usepackage{paralist}
\usepackage{xcolor}
\usepackage{hyperref}
\usepackage{pgfplots}
\pgfplotsset{compat=1.18}
\usepackage[top=0.7in, left=0.673in, right=0.7in, bottom=1.013in]{geometry}
\def\BibTeX{{\rm B\kern-.05em{\sc i\kern-.025em b}\kern-.08em
    T\kern-.1667em\lower.7ex\hbox{E}\kern-.125emX}}
\definecolor{cobalt}{rgb}{0.0, 0.28, 0.67}
\definecolor{britishracinggreen}{rgb}{0.0, 0.26, 0.15}
\definecolor{burntorange}{rgb}{0.8, 0.33, 0.0}
\definecolor{lightgreen}{rgb}{0.56, 0.93, 0.56}
\definecolor{lightsalmon}{rgb}{1.0, 0.63, 0.48}
\definecolor{lightblue}{rgb}{0.68, 0.85, 0.9}
\definecolor{lightseagreen}{rgb}{0.13, 0.7, 0.67}
\definecolor{lightskyblue}{rgb}{0.53, 0.81, 0.98}
\definecolor{lightcarminepink}{rgb}{0.9, 0.4, 0.38}
\definecolor{goldenpoppy}{rgb}{0.99, 0.76, 0.0}
\definecolor{blue}{rgb}{0,0,0}
\definecolor{grannysmithapple}{rgb}{0.61, 0.86, 0.57}
\definecolor{cornflowerblue}{rgb}{0.40,0.60,1.00}
\usepackage{tikz}
\usetikzlibrary{external}

\usetikzlibrary{shapes.geometric, arrows}
\tikzstyle{S1} = [rectangle, rounded corners, minimum width=2cm, minimum height=1.5cm,text centered,text width=2cm, draw=black, fill=cornflowerblue!50]
\tikzstyle{S2} = [rectangle, rounded corners, minimum width=2cm, minimum height=1.5cm,text centered,text width=2cm, draw=black, fill=grannysmithapple!80]
\tikzstyle{Proposed_Framework} = [rectangle, rounded corners, minimum width=2cm, minimum height=1.5cm,text centered,text width=2cm, draw=black, fill=grannysmithapple!80]
\usepackage{bm}
\usepackage{amsmath,amssymb,amsfonts}
\newcommand{\C}{\mathbb{C}}
\newcommand{\NT}{N_T}

\newcommand{\R}{\mathbb{R}}
\newcommand{\cN}{\mathcal{N}}
\newcommand{\Hhat}{\hat{\mathbf{H}}}
\newcommand{\bH}{\mathbf{H}}
\newcommand{\bx}{\mathbf{x}}
\newcommand{\by}{\mathbf{y}}
\newcommand{\bw}{\mathbf{w}}

\newcommand{\NR}{N_R}  
\newcommand{\NF}{N_F}   
\newcommand{\NS}{N_S}  
\newcommand{\bI}{\mathbf{I}}
\begin{document}

\title{Towards a Joint Task-Oriented and Generative Semantic Communication Framework for 6G Networks}


%
\author{%
Soheyb Ribouh$^{*}$, Phil Polo Ditsia Di Ngoma$^{*}$, \\
$^{*}$Univ Rouen Normandie, LITIS Laboratory UR 4108, F-76000 Rouen, France\\
\thanks{The source code is publicly available at: \url{https://github.com/philpolo/rsgen}.}
}



\maketitle

\begin{abstract}
Semantic Communication (SC) has emerged as a key enabler for 6G wireless systems by transmitting task-relevant meaning rather than raw data, thereby significantly reducing bandwidth consumption while preserving communication intent. In this work, we propose an end-to-end OFDM-based semantic communication framework that integrates a semantic encoder–decoder pipeline with a neural receiver operating over a 3GPP vehicular channel. The semantic encoder extracts the underlying meaning of a visual scene by transforming it into a graph-based representation consisting of object-level features and relational structure. At the receiver, the reconstructed scene graph is processed by a spatio-temporal graph neural network (ST-GNN)-based module for collision-risk estimation, enabling task-oriented inference. In parallel, a diffusion-based semantic decoder reconstructs the visual scene from the recovered semantics, providing dual functionality: safety prediction and image reconstruction. The proposed framework is evaluated in a MIMO configuration under varying SNR conditions. Experimental results show that it achieves up to $99.1\%$ data compression relative to pixel-domain transmission, outperforming conventional compression-based methods (JPEG and HEVC)  while preserving downstream inference performance. Furthermore, the diffusion-based reconstruction attains significantly lower fréchet inception distance (FID) scores than existing semantic communication approaches, reflecting superior semantic and perceptual fidelity.
\end{abstract}


\section{Introduction}

The next generation of wireless communication (6G) is expected to pursue the vision of achieving \textit{global connectivity} \cite{zhang2022big}. This vision is driven by the growing need to provide seamless and ubiquitous network access, extending high-speed and reliable connectivity to underserved regions while ensuring continuous service without interruption for emerging applications that demand persistent access to communication networks \cite{kalor2024wireless}. Beyond bridging geographical and digital branches, 6G is also envisioned to enable mission-critical verticals such as the automotive industry. In particular, connected automated vehicles (CAVs) represent a transformative application, where vehicles and drivers progressively transition from individual decision-making to fully collaborative autonomy \cite{ribouh2020multiple}. This evolution will enable advanced use cases such as cooperative perception and real-time traffic monitoring, while enhancing safety and efficiency in intelligent transportation systems (ITS) \cite{nguyen2022toward}. However, enabling such collaborative and safety-critical technology requires the transmission of massive amounts of sensor data. Sharing this high-volume information can result in significant bandwidth consumption, latency issues, and redundant data exchange \cite{ribouh2024semantic}. 

To address these limitations, \textit{Semantic communication (SC }) has emerged as a promising paradigm. SC leverages artificial intelligence (AI) techniques to extract and transmit only the task-relevant, meaningful information  \cite{getu2025semantic}. This will improves bandwidth efficiency,  offer an ultra-high data rate, and frees network resources for other users and services  \cite{ribouh2025large}. This paradigm shift aligns closely with the strategic standardization direction of the 3rd generation partnership project (3GPP) consortium, which has identified AI-driven approaches  and semantic communications as integral components of future wireless communication \cite{lin2023embracing}.

Early SC approaches for visual data primarily relied on latent or feature-based semantics, where Lokumarambage et al.~\cite{lokumarambage2023wireless} proposed an end-to-end image transmission framework using semantic segmentation and a GAN-based decoder, demonstrating substantial bandwidth savings compared to conventional source–channel separation. Similarly, Huang et al.~\cite{huang2022toward} introduced a reinforcement learning-assisted semantic encoder that adaptively controls the semantic rate, improving bitrate efficiency while maintaining recognizability. However, these methods operate on compact visual features rather than structured representations, which limits interpretability and downstream task reasoning.

Beyond the visual domain, SC has been extended to other modalities such as audio. Tong et al.~\cite{tong2021federated} developed a federated learning-based semantic transmission scheme for audio signals to reduce uplink communication overhead in distributed learning systems. Liang et al.~\cite{liang2025semantic} further explored semantic-aware synchronization mechanisms for the internet of sound (IoS), showing that higher-level semantic priors can reduce latency and improve temporal coherence in multimodal sensing. These studies illustrate that semantic compression is modality-agnostic, but they do not address scene-level reasoning required for autonomous driving.

To improve interpretability and support structured downstream tasks, recent works have adopted graph-based semantic representations. Sun et al.~\cite{sun2024task} introduced GRACE, a scene graph-based semantic communication system for image retrieval tasks, demonstrating greater robustness than pixel-domain transmission. Similarly, Wang et al.~\cite{wang2025explicit} proposed an explicit semantic-base architecture to enhance semantic reusability for multi-task scenarios. In parallel, knowledge graph–driven semantic communication frameworks have shown that incorporating structured prior knowledge enables semantic inference and improves robustness compared to purely latent feature representations \cite{zhou2023cognitive}. These advances further highlight the potential of structured representations to bridge perception-driven semantics and symbolic reasoning. These systems, however, typically assume idealized channels and do not integrate with a realistic wireless PHY layer.

Multi-task semantic communication has also attracted interest. Zhang et al.~\cite{zhang2024unified} proposed a unified multimodal semantic transmission system capable of jointly supporting multiple perception tasks. While their system highlights the versatility of SC, it does not address vehicular communication constraints or PHY-layer robustness. In contrast, our approach focuses specifically on graph-based semantics for vehicular safety tasks and integrates them directly into the wireless transmission pipeline.

A parallel line of research has begun to investigate generative models for semantic reconstruction. Guo et al.~\cite{guo2025diffusion} demonstrated the potential of diffusion models to improve image reconstruction quality in SC by operating beyond latent-space coding. However, their method does not incorporate structured semantics such as scene graphs, and still assumes simplified channel conditions. In addition, diffusion-based SC has not yet been explored in combination with semantic reasoning tasks such as collision prediction.

From a physical-layer perspective, Liu et al.~\cite{liu2024ofdm} proposed an OFDM-based digital semantic communication system with semantic importance awareness, where subcarrier and bit allocation are optimized according to the task relevance of semantic features. Their results demonstrate that importance-aware transmission significantly improves robustness to channel impairments compared to conventional communication schemes. However, these systems still rely on feature-level semantic representations and do not support generative semantic reconstruction. 

In recent work Diao et al.~\cite{diao2025aligning} propose a framework that bridges task-oriented and reconstruction-oriented communication paradigms for edge intelligence. The authors introduce an information reshaping mechanism combined with joint source channel coding to align inference performance with signal reconstruction objectives. However, their work lacks consideration of a realistic wireless transmission setting based on standardized 3GPP channel models, and a full PHY-layer communication pipeline is not addressed.

In contrast to all prior studies, the framework proposed in this paper jointly addresses (i) structured scene-level semantics via scene graphs, (ii) task-oriented reasoning (collision prediction), (iii) high-fidelity semantic reconstruction using graph-conditioned diffusion models, and (iv) full PHY-layer integration over a 3GPP 
vehicular channel with a neural receiver. To the best of our knowledge, this is the first semantic communication framework that unifies graph-based semantics, generative reconstruction, and realistic MIMO channel modeling in a single end-to-end system.

In this work, we propose a semantic communication framework for MIMO vehicular networks that jointly enables task-oriented inference and semantic image reconstruction over a realistic wireless channel. The main contributions of this paper are summarized as follows:
\begin{itemize}

\item We introduce a graph-based semantic communication architecture for autonomous driving that transmits structured scene semantics rather than pixel-domain data. The semantic encoder converts visual scenes into scene graphs capturing object-level features and inter-object relationships, while the semantic decoder reconstructs these graphs for downstream processing. The reconstructed semantics are used for both (i) collision-risk prediction via a GCNN–LSTM pipeline and (ii) semantic scene reconstruction through a Stable-Diffusion-based generative module.

\item We integrate the proposed semantic communication pipeline into a full end-to-end MIMO wireless transmission system equipped with a neural receiver, and evaluate its performance under a 3GPP-compliant  vehicular channel model across a wide range of SNR conditions. This allows us to jointly assess semantic fidelity, task performance, and reconstruction quality under realistic channel impairments.
\end{itemize}

The remainder of this paper is organized as follows. Section~\ref{Proposed_Framework} presents the proposed semantic communication framework, including both the semantic and bit-level processing modules. Section~\ref{Imp_Exp} describes the training methodology and experimental setup. Section~\ref{Results} reports performance results and provides a comparative analysis against state-of-the-art approaches. Finally, Section~\ref{Conclusion} concludes the paper and outlines future research directions.


\section{Proposed Framework} \label{Proposed_Framework}
\begin{figure*}[ht]
    \centering
    \includegraphics[width=1.0\linewidth]{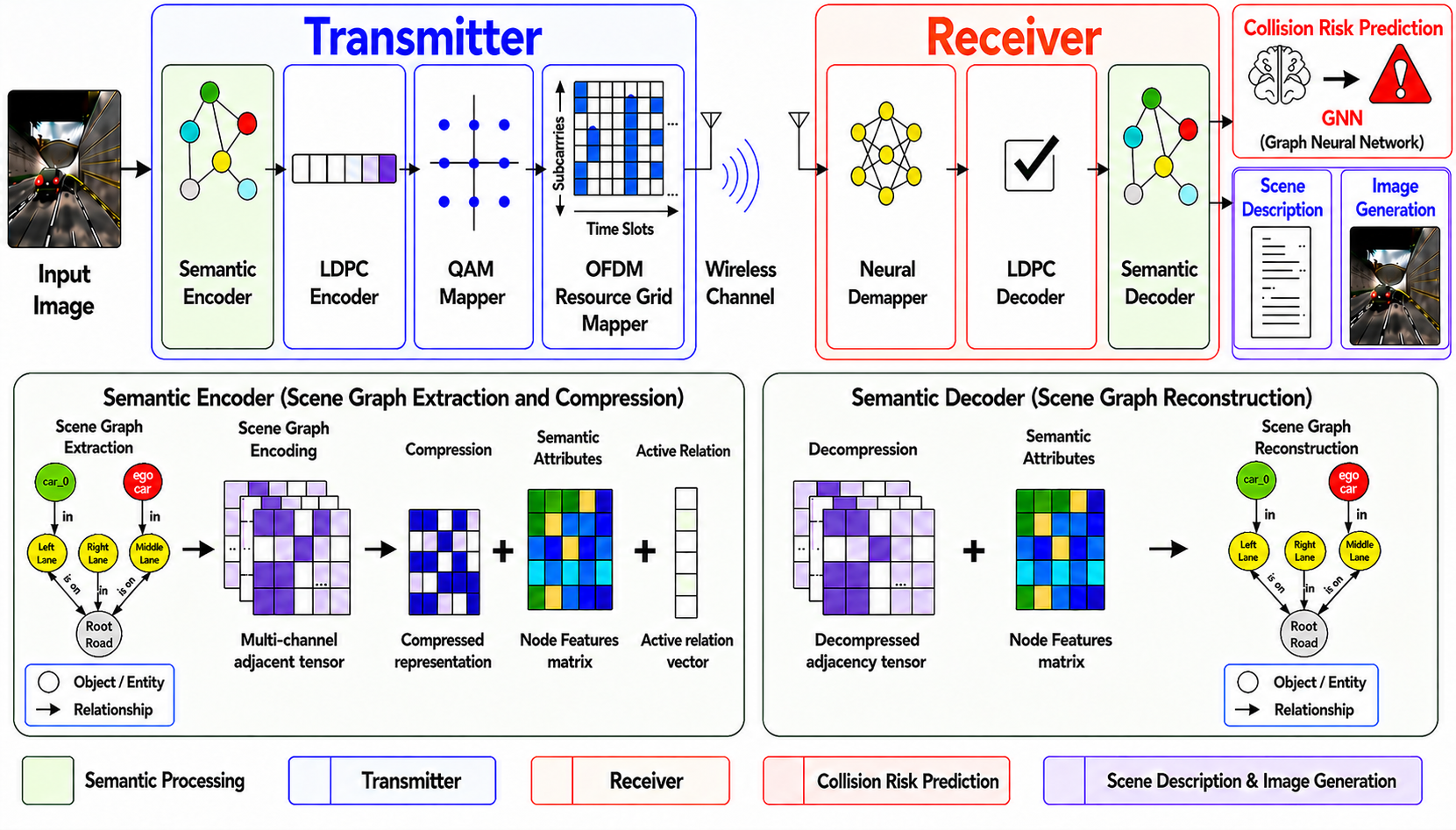}
    \caption{End-to-end semantic communication framework. The transmitter encodes a scene graph from the input image and transmits it over a wireless channel, while the receiver reconstructs the graph for collision risk prediction and scene understanding.}
    \label{fig:architecture}
\end{figure*}

The proposed semantic communication framework is illustrated in Fig. \ref{fig:architecture}. Unlike conventional communication systems that rely on signal reconstruction prior to task inference, the proposed framework directly operates on structured semantic representations, where task-relevant information is transmitted without requiring full signal reconstruction. This design avoids error propagation from reconstruction to task inference and improves overall system efficiency. The system operates across two levels:

\begin{enumerate}  
  \item \textbf{Semantic level} A graph-based semantic encoder extracts and compresses task-relevant information from the input data into a structured scene graph representation. The scene graph serves as the primary information carrier for collision prediction. At the receiver, a semantic decoder reconstructs this representation, enabling task-oriented inference (collision prediction), while a generative reconstruction module provides an auxiliary visual interpretation of the transmitted semantics through scene generation.

    \item \textbf{Bit level} A wireless communication module bridges the semantic encoder and decoder by mapping the compressed semantic representation into a bitstream, transmitting it over the physical channel, and recovering it at the receiver.  
\end{enumerate}  

By jointly integrating the semantic and bit levels, the framework enables efficient compression, robust transmission, and accurate reconstruction, thereby supporting both task-oriented communication and scene  generation. The details of each level and the components included in it are provided in the following subsections.

\subsection{Semantic level}
\subsubsection{\textbf{Semantic Encoder}} \label{SE_EN}
The semantic encoder is a key component designed to identify the most effective low-dimensional representation of the meaning conveyed by the input image. As outlined in Algorithm~\ref{alg:SE_EN}, the encoder operates through a two-step process as follows:

\paragraph{\textbf{Scene Graph Extraction}}  
The scene graph extraction process transforms the input image into a structured representation that explicitly captures the relationships among objects in the scene. It proceeds through three main steps:  
\textbf{(1) Object detection}, which identifies and localizes all entities present in the image;  
\textbf{(2) Inverse perspective mapping (IPM)}, which projects detected objects into a bird’s eye view (BEV) coordinate system to obtain spatially consistent positions with respect to the ego frame; and  
\textbf{(3) Relationship estimation}, which infers pairwise relations between objects based on their relative spatial configurations and orientations in the BEV space.  

The result of this process is a structured graph representation  
\[
\mathcal{G} = (\mathcal{V}, \mathcal{E}, \mathcal{R}),
\]
where $\mathcal{V}$ denotes the set of detected objects (nodes), $\mathcal{E}$ the set of edges representing pairwise connections, and $\mathcal{R}$ the set of relation types inferred among them.  

\paragraph{\textbf{Scene Graph Encoding and Compression}}  
The extracted scene graph encodes both the semantic meaning and the structural relationships present in the input image. This representation is expressed through three key components:  
\begin{itemize}
    \item \textbf{Node feature matrix} $\mathbf{F}$, which aggregates the attributes of all detected objects, including geometric properties, appearance features, and class embeddings. This matrix facilitates object-level reasoning and supports feature-based reconstruction.  
    \item \textbf{Adjacency tensor} $\mathbf{T}$, which consists of multiple relation-specific adjacency matrices. Each slice of $\mathbf{T}$ corresponds to a predefined relation type, where an entry $T_{ij,k}=1$ indicates that nodes $i$ and $j$ share relation $r_k$, and $T_{ij,k}=0$ otherwise.  
    \item \textbf{Active relation set} $\mathcal{R}'$, which includes only the relation types that are actually present in the current scene. This pruning step ensures compactness and reduces the overall memory footprint of the representation.  
\end{itemize}

The final encoded representation of the visual scene is therefore given by  
\[
\{\mathbf{T}, \mathbf{F}, \mathcal{R}'\},
\]
which jointly preserves the semantic attributes of individual objects, the structural topology of their interactions, and the subset of active relations that define the scene context.

The overall process of scene graph extraction and encoding is summarized in Algorithm~\ref{alg:SE_EN}, which outputs the node feature matrix $\mathbf{F}$, the adjacency tensor $\mathbf{T}$, and the active relation set $\mathbf{R}'$ as the semantic representation of the visual scene.  

\begin{algorithm}[t]
\caption{Semantic Encoder}
\label{alg:SE_EN}
\begin{algorithmic}[1]
\State \textbf{Input:} Image $\mathbf{I}$; object detector $\mathcal{D}$; IPM parameters $\{\mathbf{K}, \mathbf{E}, \mathbf{H}_{\text{ipm}}\}$; relation set $\mathcal{R}=\{r_1,\dots,r_{|\mathcal{R}|}\}$ with predicates $\{\pi_{r}\}$; feature extractor $\varphi$; confidence threshold $\tau$
\State \textbf{Output:} Adjacency tensor $\mathbf{T}\in \{0,1\}^{N\times N\times |\mathcal{R}'|}$; node feature matrix $\mathbf{F}\in \mathbb{R}^{N\times d}$; active relation set $\mathcal{R}'\subseteq\mathcal{R}$
\vspace{5pt}

\State \textbf{(1) Object Detection:}
\Statex \hspace{1em} Detect objects $\mathcal{O}=\{(b_i,c_i,s_i)\}_{i=1}^{\tilde N}\!\leftarrow\!\mathcal{D}(\mathbf{I})$ and keep detections with $s_i\ge\tau$.
\Statex \hspace{1em} Define node set $\mathcal{V}=\{v_1,\dots,v_N\}$ where $N=\lvert\mathcal{O}\rvert$.
\vspace{3pt}

\State \textbf{(2) BEV Projection:}
\Statex \hspace{1em} For each object $v_i$, compute BEV pose $\mathbf{p}_i=(x_i,y_i,\theta_i)$ using $\{\mathbf{K},\mathbf{E},\mathbf{H}_{\text{ipm}}\}$.
\Statex \hspace{1em} Set ego/reference pose $\mathbf{p}_{\text{ego}}=(0,0,0)$ in the BEV frame.
\vspace{3pt}

\State \textbf{(3) Relation Estimation:}
\Statex \hspace{1em} Initialize empty edge set $\mathcal{E}\leftarrow\emptyset$ and adjacency tensor $\mathbf{T}\leftarrow\mathbf{0}\in\{0,1\}^{N\times N\times|\mathcal{R}|}$.
\For{$k=1$ \textbf{to} $|\mathcal{R}|$} \Comment{Each relation type $r_k$}
    \ForAll{ordered pairs $(i,j),\, i\neq j$}
        \State Compute relative descriptor $\Delta_{ij}=[x_j-x_i,\,y_j-y_i,\,\theta_j-\theta_i,\,\text{dist}_{ij},\,\text{bearing}_{ij}]$
        \If{$\pi_{r_k}(\mathbf{p}_{\text{ego}}, \mathbf{p}_i, \mathbf{p}_j, \Delta_{ij})=\texttt{true}$}
            \State $\mathbf{T}[i,j,k]\leftarrow 1$; add edge $(v_i,v_j,r_k)$ to $\mathcal{E}$
        \EndIf
    \EndFor
\EndFor
\vspace{3pt}

\State \textbf{(4) Scene Graph Assembly:}
\Statex \hspace{1em} Construct the scene graph $\mathcal{G}=(\mathcal{V},\mathcal{E},\mathcal{R})$ representing all detected entities and relations.
\vspace{3pt}

\State \textbf{(5) Scene Graph Encoding and Compression:}
\For{each node $v_i\in\mathcal{V}$}
    \State Compute geometric attributes $\mathbf{g}_i=[x_i,y_i,w_i,h_i,\theta_i]$
    \State Extract appearance embedding $\mathbf{a}_i \leftarrow \varphi(\mathbf{I}\!\!\restriction_{b_i})$
    \State Obtain class embedding $\mathbf{e}_i \leftarrow \text{Embed}(c_i)$
    \State Assemble node feature $\mathbf{f}_i \leftarrow \text{concat}(\mathbf{g}_i,\mathbf{a}_i,\mathbf{e}_i)$
\EndFor
\State Stack features into $\mathbf{F}=[\mathbf{f}_1;\dots;\mathbf{f}_N]\in\mathbb{R}^{N\times d}$.
\State Prune inactive relations: $\mathcal{K}=\{k:\sum_{i,j}\mathbf{T}[i,j,k]>0\}$; set $\mathcal{R}'=\{r_k\}_{k\in\mathcal{K}}$ and compress $\mathbf{T}$ to $|\mathcal{R}'|$ slices.
\vspace{3pt}

\State \textbf{Return:} $\{\mathbf{T}, \mathbf{F}, \mathcal{R}'\}$
\end{algorithmic}
\end{algorithm}


\subsubsection{\textbf{Semantic Decoder}}  \label{Se_DE}
As outlined in Algorithm~\ref{alg:SE_DE}, the semantic decoder reverses the process performed by the semantic encoder. It reconstructs the original scene graph from the received compact representation by first decompressing the adjacency tensor and then rebuilding the complete relational structure of the scene. This reconstructed graph serves as the input to two parallel functional branches within the decoder:  

\begin{itemize}  
    \item \textbf{Task-oriented branch:} The reconstructed graph is fed into a neural network model to perform collision prediction.  

   \item \textbf{Image regeneration branch:} The reconstructed graph is converted into a structured textual description, which is used as conditioning input to a Stable Diffusion model to regenerate the transmitted image. The textual representation enables bridging scene graphs and vision–language models, supporting a modular design and avoiding the need to retrain generative models.  The image generation module thus serves as an auxiliary component for visualization. 
\end{itemize}  

By integrating both task-oriented inference and semantic image generation, the semantic decoder provides a comprehensive output that enables efficient communication while preserving semantic fidelity.

\paragraph{\textbf{Decompression}}  
The decompression stage restores the full adjacency tensor by reintroducing zero matrices corresponding to inactive relations that were pruned during encoding.  
Given the received compressed tensor $\mathbf{T} \in \{0,1\}^{N\times N\times |\mathcal{R}'|}$ and the active relation set $\mathcal{R}' \subseteq \mathcal{R}$, the decoder initializes an empty tensor $\widehat{\mathbf{T}} \in \{0,1\}^{N\times N\times |\mathcal{R}|}$.  
For each active relation $r_k \in \mathcal{R}'$, its slice is placed back at the appropriate position corresponding to its global index within the full relation set $\mathcal{R}$, while all missing relations remain zero-filled.  
This step ensures that the decompressed tensor $\widehat{\mathbf{T}}$ matches the dimensionality of the original encoder representation.

\paragraph{\textbf{Scene Graph Reconstruction}}  
After decompression, the decoder reconstructs the scene graph by iterating over all relation types $r_m \in \mathcal{R}$ and identifying active links between node pairs.  
For each pair $(v_i, v_j)$ such that $\widehat{\mathbf{T}}[i,j,m] = 1$, a relational triplet $(v_i, r_m, v_j)$ is added to the edge set $\widehat{\mathcal{E}}$.  
The node set $\mathcal{V} = \{v_1,\dots,v_N\}$ is obtained directly from the received feature matrix $\mathbf{F}$, and the final reconstructed scene graph is given by  
\[
\mathcal{G} = (\mathcal{V}, \widehat{\mathcal{E}}, \mathcal{R}),
\]
where $\widehat{\mathcal{E}}$ denotes the set of recovered edges and $\mathcal{R}$ the complete relation set.

\begin{algorithm}[t]
\caption{Semantic Decoder}
\label{alg:SE_DE}
\begin{algorithmic}[1]
\State \textbf{Input:} Compressed adjacency tensor $\mathbf{T}\in\{0,1\}^{N\times N\times |\mathcal{R}'|}$; node feature matrix $\mathbf{F}\in\mathbb{R}^{N\times d}$; active relation set $\mathcal{R}'\subseteq\mathcal{R}$; 
\State \textbf{Output:} Reconstructed scene graph $\mathcal{G}=(\mathcal{V},\widehat{\mathcal{E}},\mathcal{R})$
\vspace{4pt}

\State \textbf{(1) Decompression (Zero-Fill Missing Relations):}
\Statex \hspace{1em} Initialize $\widehat{\mathbf{T}}\leftarrow \mathbf{0}\in\{0,1\}^{N\times N\times |\mathcal{R}|}$
\For{$k=1$ \textbf{to} $|\mathcal{R}'|$}
    \State Let $r_k\in\mathcal{R}'$ and $m\leftarrow \mathrm{idx}(r_k)$ \Comment{global index in $\mathcal{R}$}
    \State $\widehat{\mathbf{T}}[:,:,m] \leftarrow \mathbf{T}[:,:,k]$
\EndFor
\vspace{4pt}

\State \textbf{(2) Graph Reconstruction:}
\Statex \hspace{1em} Define node set $\mathcal{V}=\{v_1,\dots,v_N\}$ from rows of $\mathbf{F}$
\Statex \hspace{1em} Initialize $\widehat{\mathcal{E}}\leftarrow \emptyset$
\For{$m=1$ \textbf{to} $|\mathcal{R}|$}
    \ForAll{ordered pairs $(i,j),\, i\neq j$}
        \If{$\widehat{\mathbf{T}}[i,j,m]=1$}
            \State Add edge $(v_i,v_j,r_m)$ to $\widehat{\mathcal{E}}$
        \EndIf
    \EndFor
\EndFor
\State Construct $\mathcal{G}=(\mathcal{V},\widehat{\mathcal{E}},\mathcal{R})$
\vspace{4pt}

\State \textbf{Return:} $\mathcal{G}$
\end{algorithmic}
\end{algorithm}

Once the scene graph $\mathcal{G}$ has been reconstructed, the system proceeds to the task-oriented inference stage. As described above, it comprises two parallel branches: (1) an image regeneration module that reconstructs the visual scene, and (2) a collision prediction module that estimates the risk level based on the decoded semantic relations.
\subsubsection{\textbf{Image Reconstruction}} \label{ig}
The reconstructed scene graph $\mathcal{G}$ is used to regenerate the visual scene through a two-step process:  
\begin{inparaenum}[(i)]
    \item converting the decoded scene graph into a sequential textual description, and  
    \item using this textual sequence as input to a Stable Diffusion model for image synthesis.
\end{inparaenum}  
Each step is detailed below.

\paragraph{\textbf{Scene Graph Description}} \label{graph_To_text}
The reconstructed graph $\mathcal{G}$ is first converted into a textual representation that captures the semantics of all recovered relations.  
This is achieved by constructing descriptive sentences from the triplets $(v_i, r_m, v_j)$, where $v_i, v_j \in \mathcal{V}$ are nodes and $r_m \in \mathcal{R}$ denotes the relation connecting them.  
Here, the initial node $v_i$ acts as the subject, the relation $r_m$ serves as the predicate, and the final node $v_j$ functions as the direct object   
All generated sentences are concatenated to form a coherent textual description summarizing the relational structure and object interactions encoded in $\mathcal{G}$.

\paragraph{\textbf{Image Generation}}  
The textual description derived from $\mathcal{G}$ is then processed by the Stable Diffusion model \cite{rombach2022high}, which synthesizes an image consistent with the described scene.  
Stable Diffusion operates through a two-phase process: a \textit{forward diffusion} phase, where Gaussian noise is progressively added to an image until complete corruption, followed by a \textit{reverse denoising} phase that iteratively reconstructs the image from noise.  
The denoising process is conditioned on a text embedding situated in a joint text–image latent space, ensuring semantic alignment between the generated image and the input description.  
It can also be further guided by auxiliary conditioning inputs such as semantic maps or sketches that refine the visual layout.  
Unlike conventional diffusion models that operate directly in the pixel space, Stable Diffusion performs all operations in a compressed latent space, significantly reducing computational requirements while maintaining high fidelity visual reconstruction.

\subsubsection{\textbf{Collision Prediction}} \label{collision_prediction}

The collision prediction module uses the reconstructed scene graph $\mathcal{G}$ to estimate the likelihood of collisions over time. To effectively interpret the structural and relational information encoded in $\mathcal{G}$, we employ a spatio-temporal graph neural network (ST-GNN) architecture. The spatial component is implemented using graph convolutional networks (GCNs), combined with graph pooling and readout layers to capture both node- and edge-level dependencies. The resulting output, denoted by $h_G^t$, represents a compact embedding that summarizes the semantic and spatial configuration of the scene at time $t$.

To model temporal dynamics, the sequence of scene embeddings $\{h_{G_1}, h_{G_2}, \dots, h_{G_T}\}$ is processed by a recurrent temporal network. Specifically, a long short-term memory (LSTM) architecture aggregates the current spatial representation $h_{G_t}$ with contextual information from previous frames through its hidden and cell states $(p_{t-1}, c_{t-1})$, producing an updated hidden representation $p_t$. This hidden state captures the spatiotemporal evolution of the driving environment and is subsequently passed to a multi-layer perceptron (MLP) for collision likelihood estimation.

The MLP output passes through a final activation layer, yielding two confidence scores: $\hat{y}_0$ corresponding to the probability of a safe situation, and $\hat{y}_1$ representing the probability of collision risk. The final decision for each frame is obtained through maximum-likelihood classification:
\begin{equation}
\hat{Y}_t =
\begin{cases}
1, & \text{if } \hat{y}_1 \geq \hat{y}_0 \ (\text{Collision Risk}) \\
0, & \text{otherwise} \ (\text{Safe})
\end{cases}
\end{equation}
By combining spatial reasoning via GNNs with temporal reasoning through LSTMs, this hierarchical model enables robust and context-aware prediction of collision risks in dynamic driving environments.

\subsection{Bit Level} \label{bit_level}
At this stage, the input to the bit-level processing chain is the  semantic representation of the visual scene, as detailed in Section~\ref{SE_EN}. It is given by
\{$\mathbf{T}, \mathbf{F}, \mathcal{R}'$\}.\\
These semantic structures are serialized and quantized into a binary sequence $\mathbf{b}$, compatible with the channel coding stage.

\subsubsection{\textbf{LDPC Encoding}}  
The binary stream $\mathbf{b}$ is processed by a low-density parity-check (LDPC) encoder that adds parity bits to enhance robustness against channel impairments. The encoded codeword is defined as
\begin{equation}
\mathbf{c} = \mathbf{G}_{\text{LDPC}}\mathbf{b},
\end{equation}
\noindent
where $\mathbf{G}_{\text{LDPC}}$ is the LDPC generator matrix. 

\subsubsection{\textbf{ Modulation}}  
The encoded bits are subsequently mapped to complex baseband symbols using quadrature amplitude modulation (M-QAM):
\begin{equation}
    x_{n_F,n_S,n_T}
      = \mathcal{M}\!\left(c_{n_F,n_S,n_T}\right)
      = a_{n_F,n_S,n_T} + j\,b_{n_F,n_S,n_T}
      \;\in\; \mathcal{X}_{M},
    \label{eq:qam}
\end{equation}
\noindent
where $\mathcal{M}(\cdot)$ is the modulation mapping function, $a_{n_F,n_S,n_T}, b_{n_F,n_S,n_T} \in \ \mathbb{R}$ are the in-phase~(I) and
quadrature~(Q) components, $\mathcal{X}_{M}$ is the $M$-QAM constellation alphabet, $N_F$ is the number of subcarriers, $N_S$ is the number of OFDM symbols per slot, $N_T$ is the number of transmit antennas, and $(k, \ell, n_T)$  with $k \in \{1,\ldots,N_F\}$, $\ell \in \{1,\ldots,N_S\}$, $n_T \in \{1,\ldots,N_T\}$ index subcarrier, OFDM symbol, and spatial layer respectively.

\subsubsection{\textbf{OFDM Resource Grid}}
For each spatial layer $n_T$, the modulated symbols are arranged into the OFDM resource grid (RG) as follows:
\begin{equation}
    \mathbf{X}_{n_T}
    = \begin{bmatrix}
        x_{1,1,n_T}   & \cdots & x_{1,N_S,n_T}   \\
        \vdots         & \ddots & \vdots            \\
        x_{N,1,n_T}   & \cdots & x_{N,N_S,n_T}
      \end{bmatrix}
    \in \C^{N \times N_S},
    \label{eq:rg}
\end{equation}
Where rows correspond to subcarriers $k$ and columns to OFDM symbols $\ell$.

\subsubsection{\textbf{OFDM Signal Generation}}  
The modulated symbols are organized into the OFDM resource grid together with pilot symbols for channel estimation. The time-domain OFDM signal for spatial layer $n_T$ and OFDM symbol $\ell$ is produced by the $\NF$-point inverse fast Fourier transform (IFFT):
\begin{equation}
    x_{n_T,\ell}(l)
      = \frac{1}{\sqrt{\NF}}
        \sum_{k=0}^{\NF-1}
        x_{k,\ell,n_T}\,
        e^{\,j2\pi k l / \NF},
        \quad l = 0,\ldots,\NF-1,
    \label{eq:ifft}
\end{equation}
followed by the insertion of a cyclic prefix (CP) of length $N_{\rm CP}$ chosen to exceed the maximum channel delay spread, ensuring inter-symbol-interference-free reception.

The received signal vector at subcarrier $k$ and OFDM symbol $\ell$ is:
\begin{equation}
    \by_{k,\ell}
      = \bH_{k,\ell}\,\bx_{k,\ell} + \bw_{k,\ell},
    \label{eq:channel}
\end{equation}
where $\bx_{k,\ell} = [x_{k,\ell,1},\ldots,x_{k,\ell,\NT}]^T \in \C^{\NT}$ is the vector of transmitted symbols across $\NT$ layers,
$\bH_{k,\ell} \in \C^{\NR \times \NT}$ is the MIMO channel matrix ($\NR$ denoting the number of receive antennas), $\bw_{k,\ell} \sim \cN(\mathbf{0},\sigma^2\bI_{\NR})$ is the additive white Gaussian noise (AWGN) with per-antenna power $\sigma^2$, and $\by_{k,\ell} \in \C^{\NR}$ is the received signal vector.

\subsubsection{\textbf{Neural Demaper: Joint CE, Equalization, and Demapping}}  
\paragraph{Bootstrap LS Channel Estimate}

A per-pilot least-squares (LS) estimation of the channel coefficient for layer $n_T$ is computed at every pilot position :

\begin{equation}
    \hat{h}_{k,\ell,n_T}
      = \frac{p^{*}_{k,\ell,n_T}\;\cdot\;y_{k,\ell}}
             {\bigl|p_{k,\ell,n_T}\bigr|^{2}},
    \label{eq:ls}
\end{equation}
where $p_{k,\ell,n_T} \in \C$ is the known pilot symbol transmitted by layer $n_T$, and $(\cdot)^{*}$ denotes complex conjugation. Linear interpolation over the time-frequency grid yields estimates, giving the per-layer bootstrap estimate $\Hhat_{n_T} \in \C^{\NF \times \NS \times \NR}$.

\paragraph{Neural Network Demaper}
The neural network $f_{\bm{\theta}}$ , whose architecture combines residual blocks with separable convolutional layers and a GNN component to capture multi-antenna spatial dependencies, jointly performs channel estimation, equalization, and soft demapping in a single forward pass. It maps the received resource grid $\mathbf{Y}$, the pilot positions $\{\mathcal{P}_{n_T}\}$, the bootstrap LS estimates
$\{\Hhat_{n_T}\}$, to a tensor of soft
log-likelihood ratios (LLRs):
\begin{equation}
    \bm{\ell}
      = f_{\bm{\theta}}\!\left(
          \mathbf{Y},\;
          \bigl\{\mathcal{P}_{n_T}\bigr\}_{n_T=1}^{\NT},\;
          \bigl\{\Hhat_{n_T}\bigr\}_{n_T=1}^{\NT},\;
          N_0
        \right),
    \label{eq:neural_rx}
\end{equation}
where $\bm{\ell} \in \R^{\NF \times \NS \times \NT \times m}$ contains  soft LLR values per layer. The parameters $\bm{\theta}$ are fixed (pretrained) following the architecture of \cite{cammerer2023neural}.

\subsubsection{\textbf{LDPC Decoding}}  
The LLR tensor $\bm{\ell}$ from~\eqref{eq:neural_rx} is  processed by the LDPC belief-propagation (BP) decoder to recover the transmitted bits. After $T$ iterations, the a-posteriori LLR for bit $i$ is computed as 
\begin{equation}
L_i^{\text{out}} = L_i + \sum_{j \in \mathcal{N}(i)} m_{j \to i}^{(T)},
\end{equation}
\noindent
where $L_i$ is the input LLR, $\mathcal{N}(i)$ represents the set of check nodes connected to variable node $i$, and $m_{j \to i}^{(T)}$ is the final message from check node $j$ to variable node $i$. The hard-decision rule yields the reconstructed bit:
\begin{equation}
\hat{b}_i =
\begin{cases}
0, & L_i^{\text{out}} \geq 0, \\
1, & L_i^{\text{out}} < 0,
\end{cases}
\end{equation}
\noindent
producing the final decoded bitstream $\hat{\mathbf{b}}$, which can be restructured to recover the original semantic representation $\{\hat{\mathbf{T}}, \hat{\mathbf{F}}, \hat{\mathcal{R}}'\}$ at the receiver.

\section{Implementation and Experimental Setup} \label{Imp_Exp}
This section presents the implementation details of the proposed end-to-end semantic wireless communication framework. We begin by describing the semantic-level processing pipeline, which includes both the semantic encoder and semantic decoder. At the receiver side, the semantic decoder reconstructs the transmitted scene graph and then performs task-oriented inference, instantiated either as image synthesis (via Stable Diffusion) or collision prediction. We then introduce the bit-level transmission module, responsible for LDPC encoding, modulation and OFDM-based delivery of the compressed semantic representation over the wireless channel. All experiments are conducted on a server equipped with an NVIDIA Tesla V100-PCIE-32GB GPU.

\subsection{Semantic-Level implementation} \label{semantic_level}
The semantic-level pipeline consists of two components:

\subsubsection{\textbf{Semantic Encoder}}
For semantic encoding, we employ the RoadScene2Vec framework \cite{malawade2022roadscene2vec}, which extracts a structured scene graph from each input image. It is implemented in Python using PyTorch as the deep learning backend and relies on Detectron2 for object detection. In particular, the object detection stage uses a Faster R-CNN model with a ResNet-50 backbone and feature pyramid network (FPN) to identify dynamic road users (vehicles, pedestrians, cyclists) and static infrastructure elements (traffic signs, poles, lane markings). These detected objects are embedded and linked through a geometric reasoning module that computes pairwise spatial relations based on inter-object distance and orientation cues, which yields a raw serialized graph. This raw output is further compressed, where adjacency slices corresponding to inactive relations are removed, retaining only non-zero adjacency matrices associated with active object-to-object interactions. The node feature matrix $\mathbf{F}$ is preserved, and the relation set $\mathcal{R}'$ is pruned to include only relations corresponding to the remaining active adjacency slices.

\subsubsection{\textbf{Semantic Decoder}}
The semantic decoder reconstructs the scene graph from the received semantic tuple $\{\hat{\mathbf{T}}, \hat{\mathbf{F}}, \hat{\mathcal{R}}'\}$ and subsequently performs two downstream branches: (i) image generation for semantic scene reconstruction, and (ii) collision prediction for safety assessment, as detailed below.

\textbf{(a) Dataset preparation :}
The training of both decoder branches relies on the 1043-syn dataset \cite{c0z9-1p30-21}, which consists of 1,044 driving sequences comprising a total of 50,699 images generated using the CARLA simulator \cite{dosovitskiy2017carla}. Each sequence includes rich spatial and temporal interaction cues between road users and static infrastructure, enabling supervision for both semantic reconstruction (via image synthesis) and hazard reasoning (via collision prediction).

\textbf{(b) Collision-Prediction Branch :}
For collision-risk estimation, the reconstructed scene graph embeddings are processed by a graph-based temporal model(described in section \ref{collision_prediction}) that predicts the collision likelihood for each frame. The model is trained using the Adam optimizer with a learning rate of $5 \times 10^{-5}$ and a batch size of 16 sequences, enabling temporal reasoning over safety-critical interactions.

\textbf{(c) Image-Generation Branch :}
For semantic reconstruction through visual synthesis, Stable Diffusion 1.4 \cite{rombach2022high} is used to regenerate images from textual descriptions derived from the recovered graph. The dataset is first converted into text–image pairs as described in Section \ref{graph_To_text}. Training is performed using the AdamW optimizer with a learning rate of $5 \times 10^{-6}$ and a batch size of 5. 

To guide the model toward improving reconstruction in safety-critical objects, the model was trained  using a hybrid diffusion loss composed of a global denoising term and an object-specific penalty:
\begin{equation}
    Loss_{\text{image}} = \frac{1}{S \times M \times N} 
    \sum_{c}^{S}\sum_{m}^{M}\sum_{n}^{N} (B[c,m,n] - \hat{B}[c,m,n])^2,
\end{equation}
\begin{equation}
    Loss_{\text{car}} = \frac{1}{S \times M \times N}(I[c,m,n] - \hat{I}[c,m,n])^2,
\end{equation}
\begin{equation} \label{eq:10}
    Loss = Loss_{\text{image}} + k \times Loss_{\text{car}},
\end{equation}
where $B$ and $\hat{B}$ denote the noise added and predicted across the full image, while $I$ and $\hat{I}$ denote the corresponding noise restricted to car pixels. The weighting factor $k$ increases the contribution of errors in object-level semantics, specifically for vehicles. 

This composite loss combines a standard diffusion objective with an object-specific correction term, which conditions the denoising process on vehicle pixels. As a result, the model prioritizes the accurate reconstruction of dynamic road users (i.e., the most safety-critical semantic entities) improving task-aligned semantic fidelity.

\begin{figure}[bt!]
    \centering
    \begin{subfigure}{1.0\linewidth}
        \includegraphics[width=1.0\linewidth]{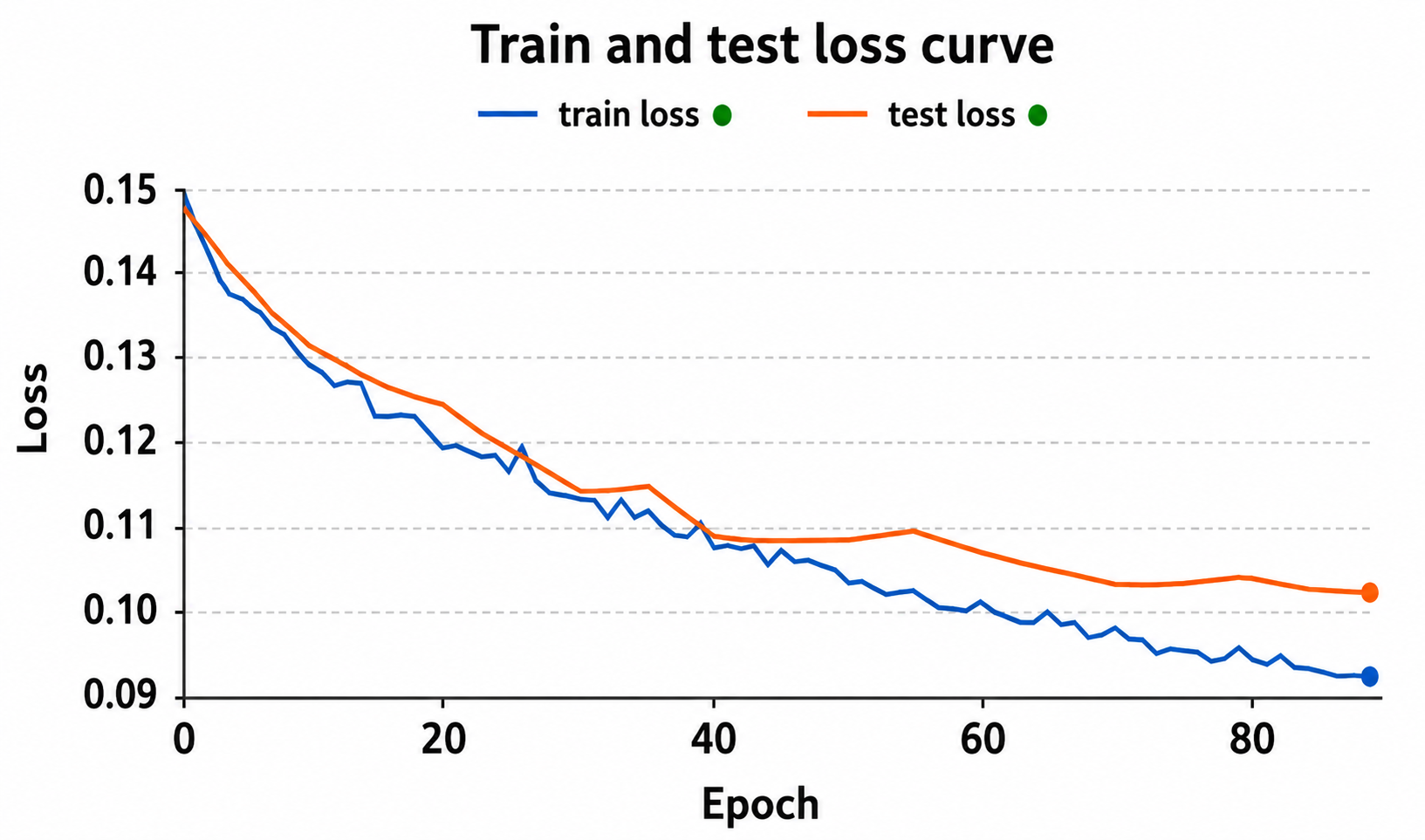}
        \caption{Global loss curve.}
    \end{subfigure}
    \begin{subfigure}{1.0\linewidth}
        \includegraphics[width=1.0\linewidth]{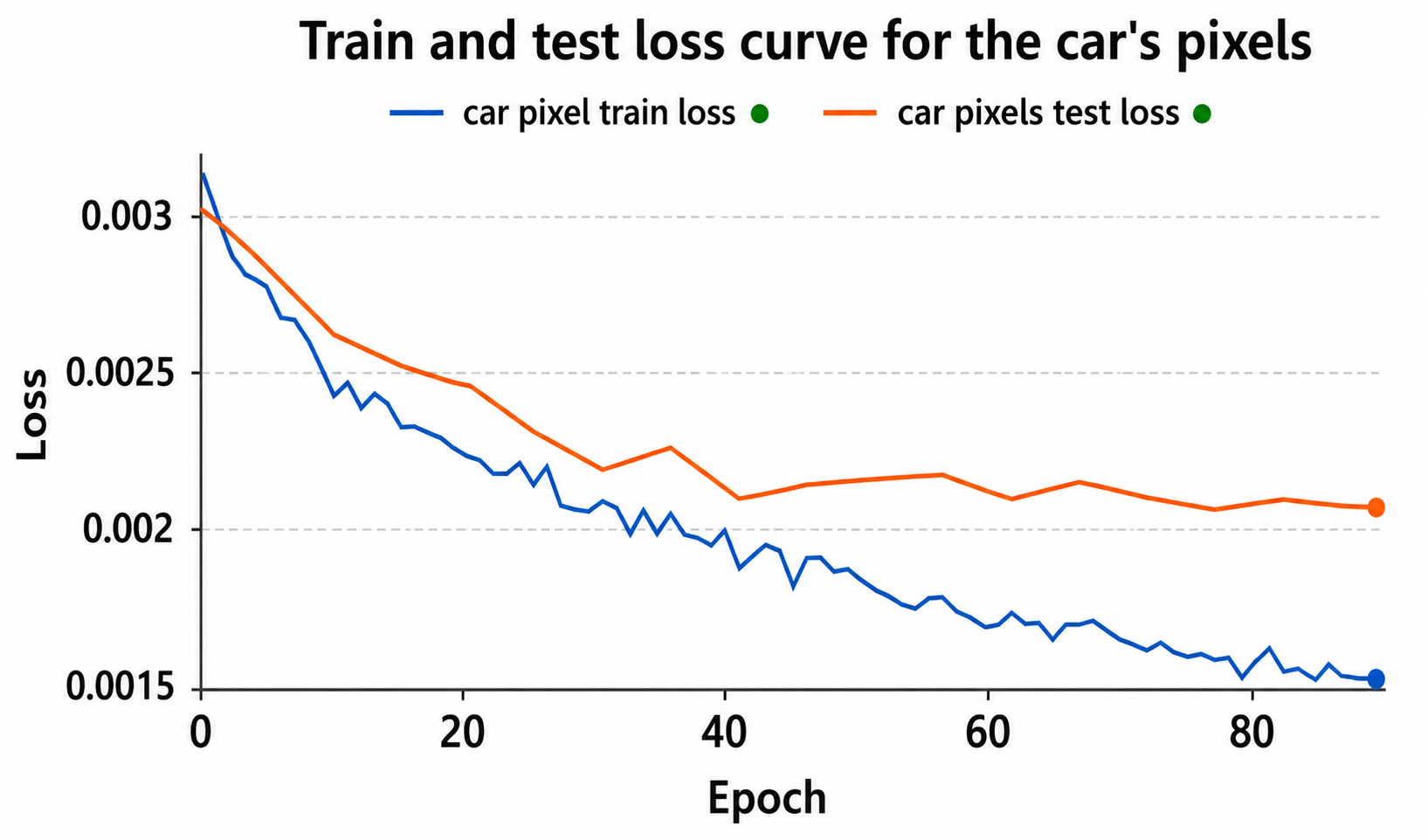}
        \caption{Car's pixels loss curve}
    \end{subfigure}
    \caption{Training loss trend over epochs.}
    \label{fig:sd_loss}
\end{figure}

The evolution of the loss during training is shown in Fig.~\ref{fig:sd_loss}, illustrating the model’s improved capability to reconstruct the transmitted images.

\subsection{Bit-Level Implementation}
The bit-level transmission module is implemented using the Sionna framework \cite{hoydis2022sionna}, which provides a differentiable and 3GPP-compliant physical layer simulation environment tailored for 6G wireless communication research. We consider an OFDM-based transmission setup consisting of $132$ subcarriers spaced at $240$ kHz. Each frame contains $14$ OFDM symbols, and a Kronecker pilot arrangement is employed to facilitate accurate channel estimation. Data symbols are mapped using $64$-QAM modulation.

A $2 \times 4$ MIMO configuration is adopted, where the semantic bitstream is transmitted through two transmit antennas over a 3GPP vehicular channel model.
For robust generalization to diverse channel conditions, the neural demapper architecture described in \cite{cammerer2023neural} is trained on a dataset of 12 million samples using the Adam optimizer with a learning rate of $10^{-3}$. 

This implementation leverages end-to-end optimization of the semantic communication chain within a realistic 6G physical-layer environment, where the semantic-level and bit-level are jointly integrated into a unified differentiable pipeline.


\section{Results and Discussion} \label{Results}

In this section, we present a comprehensive performance evaluation of the proposed end-to-end semantic communication framework. First, we quantify the compression ratio achieved at the semantic layer. Second, we evaluate the effectiveness of the collision prediction module using task-oriented performance metrics. Third, we assess the robustness of the communication pipeline under varying SNR conditions, where both semantic fidelity and collision prediction accuracy are analyzed as a function of channel quality. Finally, in addition to conventional compression baselines, we compare our method with recent semantic communication approaches using the fréchet inception distance (FID), which is widely adopted for evaluating generative reconstruction quality. This enables a fair comparison within the scope of semantic communications.

\subsection{Compression rate}

\begin{figure}[bt!]
    \centering
    \includegraphics[width=1\linewidth,height=0.65\linewidth]{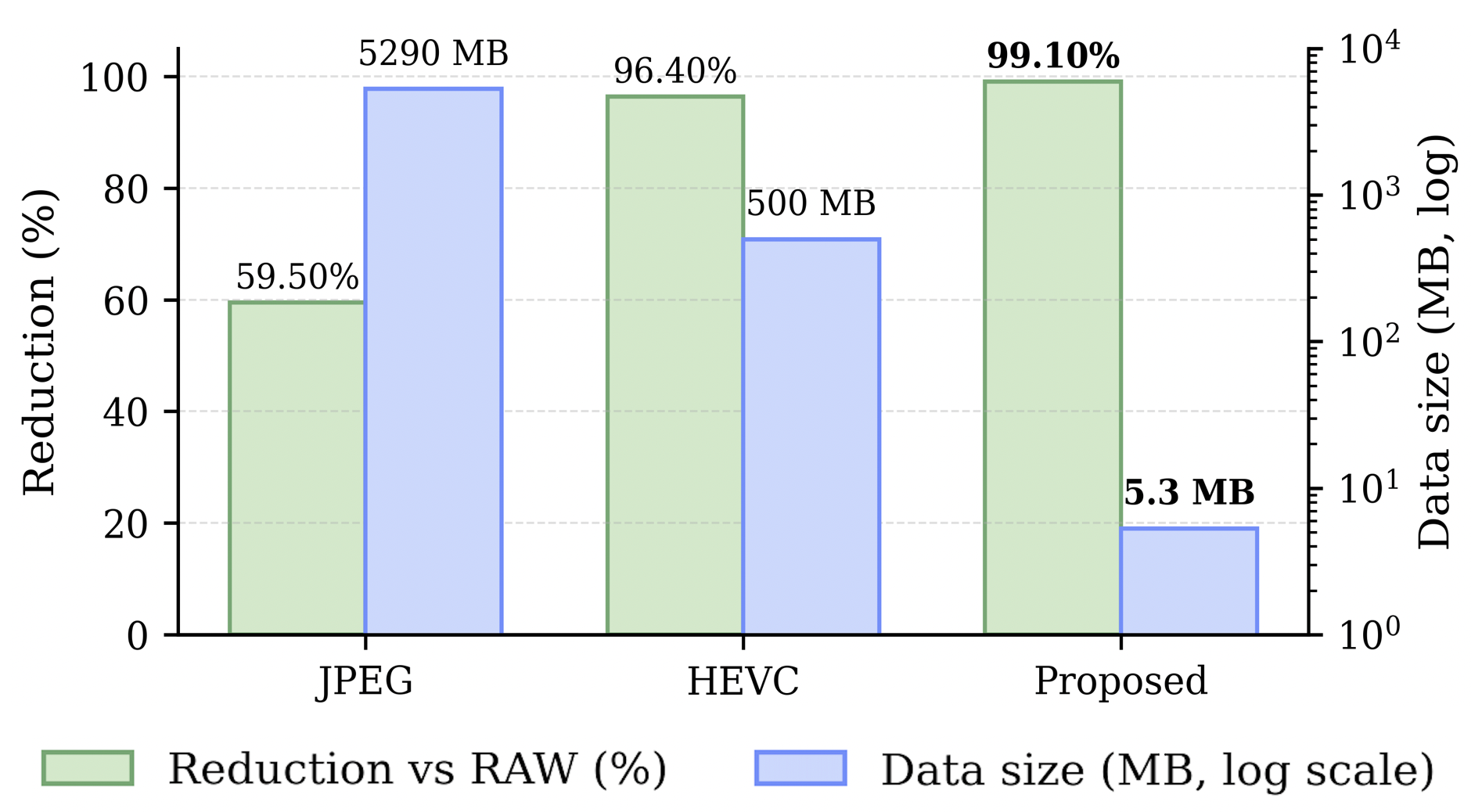}
    \caption{Compression rate}
    \label{fig:Compression_rate}
\end{figure}

We evaluate the compression capability of the proposed semantic encoder by comparing the average payload of the transmitted semantic tuple $\{\mathbf{T}, \mathbf{F}, \mathcal{R}'\}$ against conventional codecs over a validation set of 5{,}000 images. As shown in Fig.~\ref{fig:Compression_rate}, the total payloads for the full set are $5{,}290$~MB for JPEG, $500$~MB for HEVC, and $5.3$~MB for the proposed method, corresponding to reductions versus RAW of $59.5\%$, $96.4\%$, and $99.1\%$, respectively. These totals translate to per-frame averages of $\approx 1.06$~MB (JPEG), $\approx 100$~KB (HEVC), and $\approx 1.06$~KB (Proposed). Overall, the proposed semantic encoding achieves a reduction of more than three orders of magnitude relative to RAW and is $94\times$ smaller than HEVC for the same dataset, highlighting its high compression efficiency.
\subsection{Collision Prediction}

We evaluate  the collision prediction model using three standard metrics: Accuracy, Matthews Correlation Coefficient (MCC), and Area Under the ROC Curve (AUC).  The model achieves an accuracy of \textbf{0.898}, an MCC of \textbf{0.546}, and an AUC of \textbf{0.836,} indicating consistent predictive performance and strong separability between safe and hazardous driving scenes. The MCC score further confirms that the model successfully captures discriminative patterns relevant to collision likelihood.


These results demonstrate that structured semantic representations effectively capture task-relevant information for collision prediction. Importantly, this performance is achieved while transmitting significantly less data than image-based approaches, highlighting the communication efficiency of the proposed semantic communication framework.

\subsection{Communication system robustness}

\begin{figure}[bt!]
    \centering
    \includegraphics[width=1\linewidth,height=0.5\linewidth]{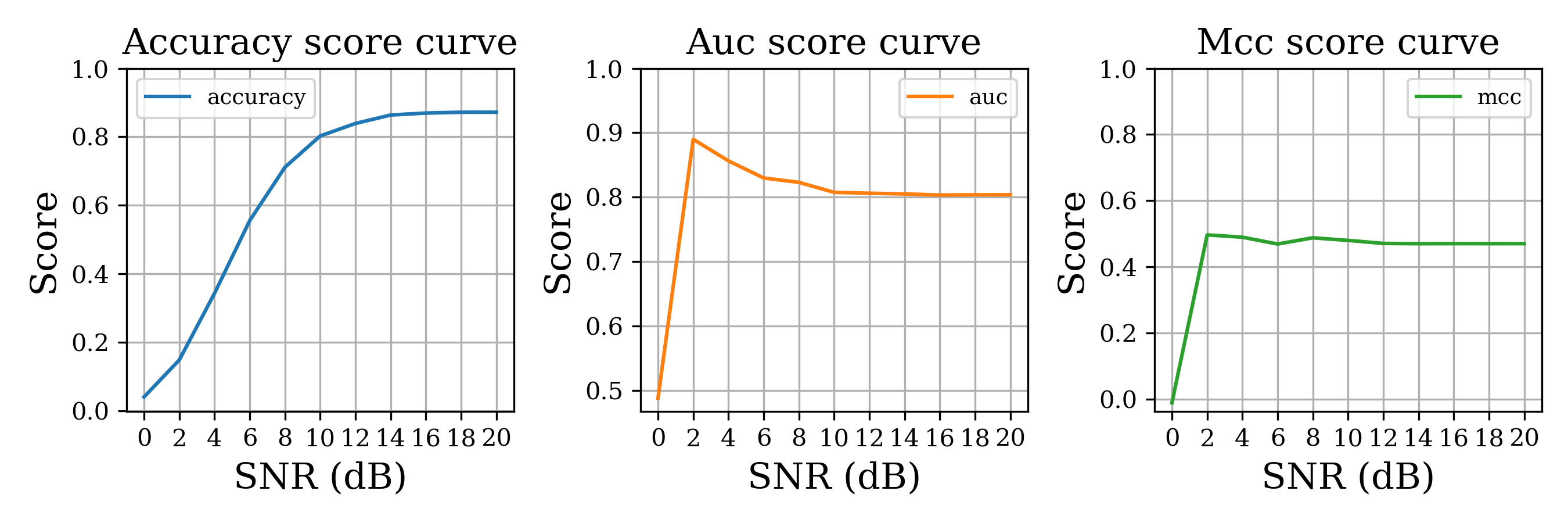}
    \caption{Collision model's performance under varying SNR}
    \label{fig:perf}
\end{figure}

To assess the communication system robustness, we evaluated the performance of the collision prediction module under varying SNR conditions, as shown in Fig.~\ref{fig:perf}. The results demonstrate that the model benefits from improved channel quality, with the accuracy score increasing steadily as the SNR rises. The AUC and MCC curves exhibit a similar trend, showing stable behavior once the SNR exceeds approximately 8 dB, indicating that the semantic representation remains sufficiently preserved for reliable decision making even in moderately noisy conditions. These results confirm that the proposed framework maintains task-level robustness under realistic channel conditions.

In addition to task-level performance, we also evaluate the system from a semantic perspective by measuring semantic fidelity under varying SNR conditions. Recent works have highlighted that semantic quality must be evaluated at the meaning level rather than at the pixel level. Getu et al.~\cite{getu2023making} discussed performance metrics for semantic and goal-oriented communication, emphasizing that conventional fidelity measures such as SSIM or PSNR are insufficient for assessing correctness in task-oriented communication. Semantic fidelity quantifies the degree to which the transmitted meaning is preserved after decoding, and can be defined as

\begin{equation}
\text{SF} = \frac{|\hat{\mathcal{S}} \cap \mathcal{S}|}{|\mathcal{S}|},
\end{equation}
where $\mathcal{S}$ denotes the set of ground-truth semantic relations extracted at the transmitter, and $\hat{\mathcal{S}}$ is the set recovered at the receiver. A value of $\text{SF}=1$ indicates perfect semantic preservation.

\begin{figure}[bt!]
    \centering
    \includegraphics[width=1\linewidth]{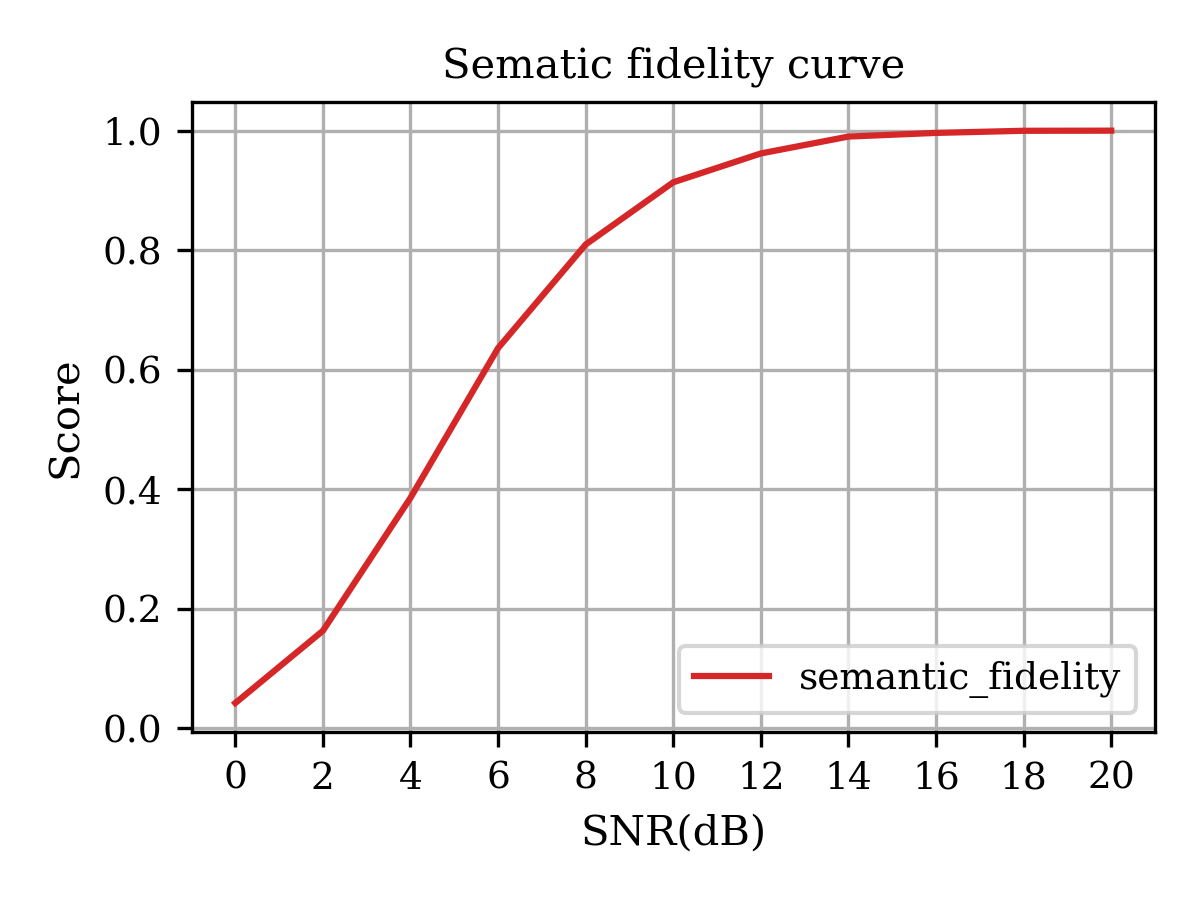}
    \caption{Semantic fidelity under varying SNR}
    \label{fig:sem_fidelity}
\end{figure}

As shown in Fig.~\ref{fig:sem_fidelity}, the semantic fidelity increases steadily with SNR. It reaches above 0.80 for SNR values greater than 8 dB. The model achieves near-perfect preservation (0.95 to 1.0) for SNR values above 12 dB, indicating that the underlying semantic structure remains intact once the channel conditions are sufficiently reliable. In particular, semantic fidelity reflects the preservation of object-level attributes and relational dependencies encoded in the scene graph. Since the collision prediction task directly relies on these spatial relationships, maintaining high semantic fidelity ensures that the task-relevant information is preserved after transmission. As a result, the proposed framework achieves stable downstream collision prediction performance even under varying channel conditions.


\subsection{Comparison to State-of-the-Art Methods}
The quality of the reconstructed images is quantitatively evaluated using the FID \cite{wu2025pragmatic}, which measures the statistical distance between the feature distributions of real and generated images. Formally, FID is defined as
\begin{equation}
\text{FID} = \|\mu_r - \mu_g\|^2 + \mathrm{Tr}\left(\Sigma_r + \Sigma_g - 2(\Sigma_r \Sigma_g)^{1/2}\right),
\end{equation}
where $(\mu_r,\Sigma_r)$ and $(\mu_g,\Sigma_g)$ denote the mean and covariance of the real and generated image feature embeddings, respectively, extracted from the Inception network. $\mathrm{Tr}$ is the trace of a matrix (sum of diagonal elements). Lower FID values indicate higher semantic and perceptual similarity between the transmitted and reference images.

\begin{figure}[t!]
    \centering
    \includegraphics[width=1.0\linewidth]{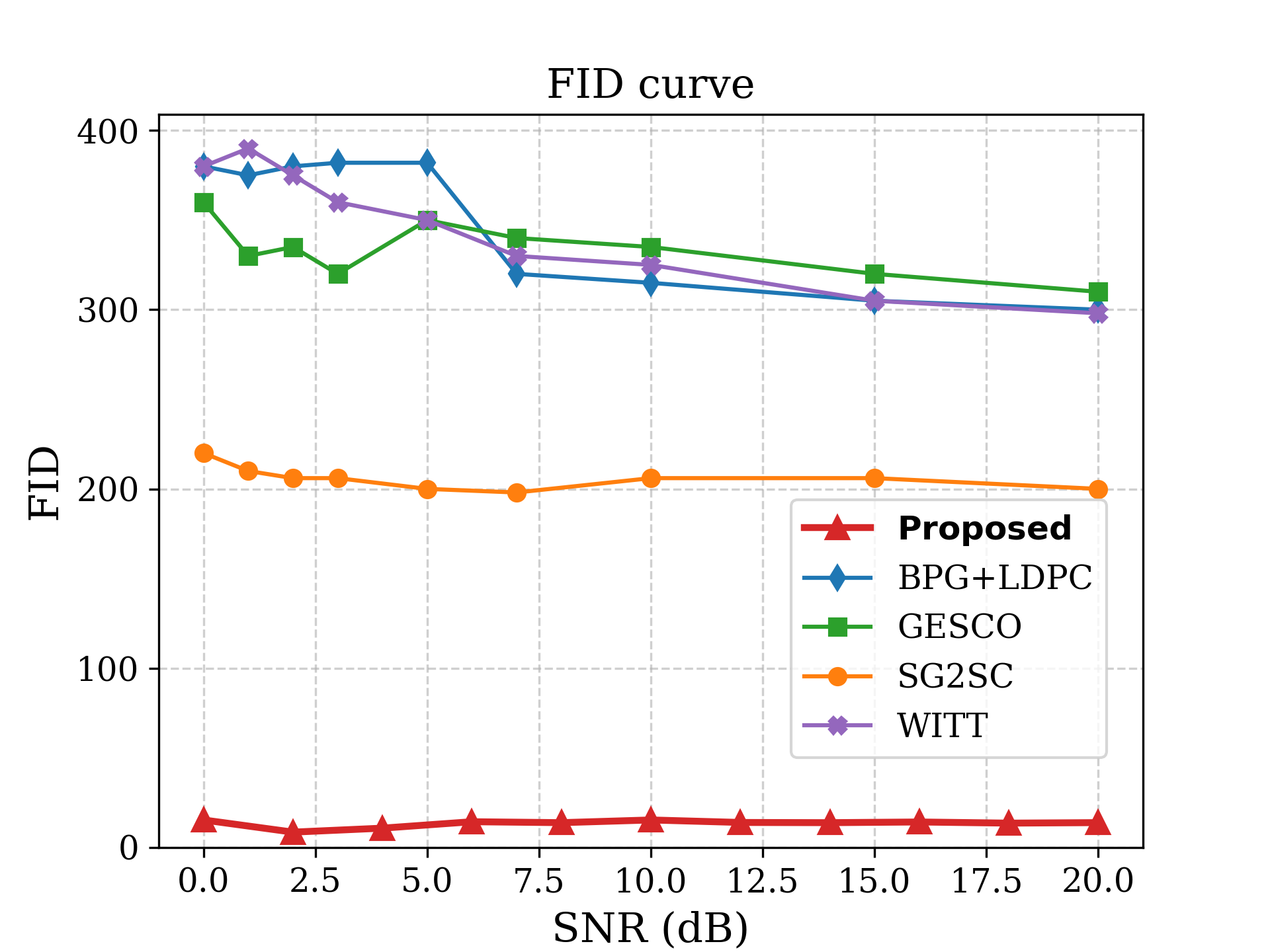}
    \caption{Fréchet Inception Distance (FID) versus SNR comparison with state-of-the-art methods}
    \label{fig:fid}
\end{figure}

Figure~\ref{fig:fid} compares our method against several representative semantic communication approaches reported in  \cite{yang2024sg2sc}, including GESCO \cite{grassucci2023generative}, WITT \cite{yang2023witt}, SG2SC \cite{yang2024sg2sc}, and the conventional BPG+LDPC pipeline. Across all tested SNR conditions, the proposed framework consistently achieves the lowest FID, ranging below 30 across diffrent SNR values, while existing semantic approaches such as SG2SC remain around 200 and BPG+LDPC and GESCO exceed 300.

This large performance gap is expected and stems from a fundamental architectural difference: unlike prior works that decode directly in pixel or latent space, our framework reconstructs the image from a structured semantic representation. The recovered scene graph provides a strong generative prior that constrains the diffusion model to a narrower manifold of semantically valid images, significantly reducing uncertainty during denoising. Furthermore, the object-level loss used during fine-tuning reinforces reconstruction quality on safety-critical objects (cars), which are dominant contributors to perceptual realism. This  highlights the effectiveness of semantic reconstruction through graph-based conditioning and confirms that the proposed model preserves both visual realism and semantic consistency.

Taken together, the semantic fidelity and FID results confirm that the proposed framework achieves dual consistency: it preserves the meaning of the transmitted scene at the semantic level while simultaneously maintaining high perceptual quality in the regenerated image. This demonstrates that the semantic encoder not only compresses task-relevant information efficiently, but also retains sufficient structure for faithful downstream reconstruction under realistic wireless channel conditions.

\subsubsection{Qualitative Visual Comparison}

\begin{figure}[t!]
    \centering
    \begin{minipage}[b]{0.48\linewidth}
        \centering
        \includegraphics[width=\linewidth,height=1\linewidth]{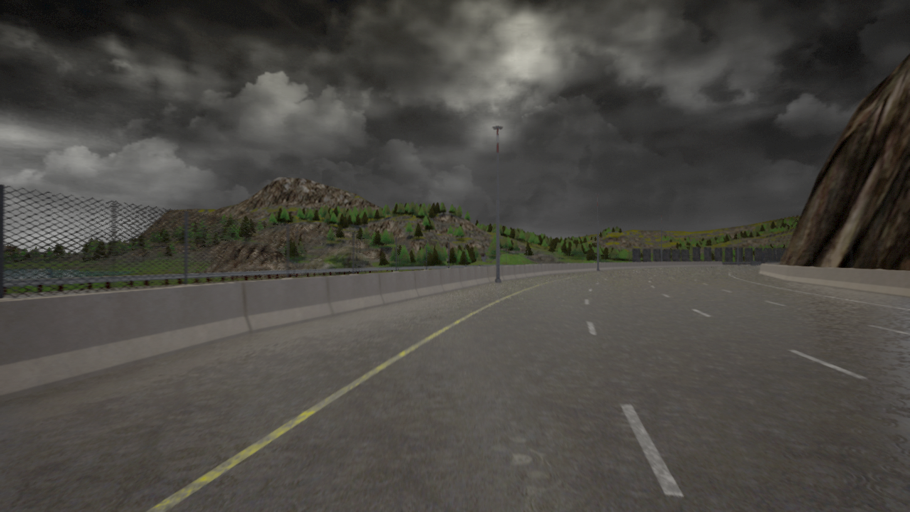}
    \end{minipage}
    \begin{minipage}[b]{0.48\linewidth}
        \centering
        \includegraphics[width=\linewidth]{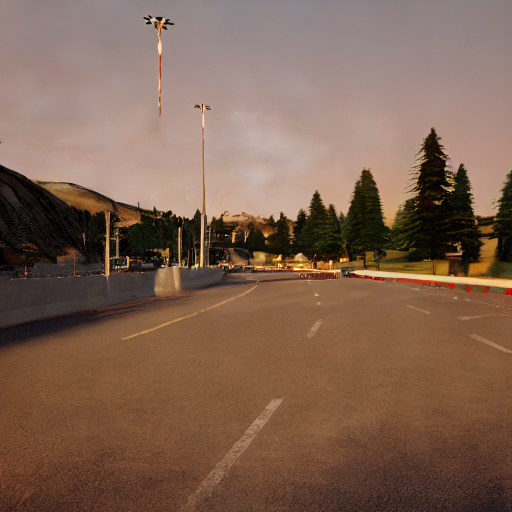}
    \end{minipage}
    
    \vspace{0.4cm} 

    \begin{minipage}[b]{0.48\linewidth}
        \centering
        \includegraphics[width=\linewidth,height=1\linewidth]{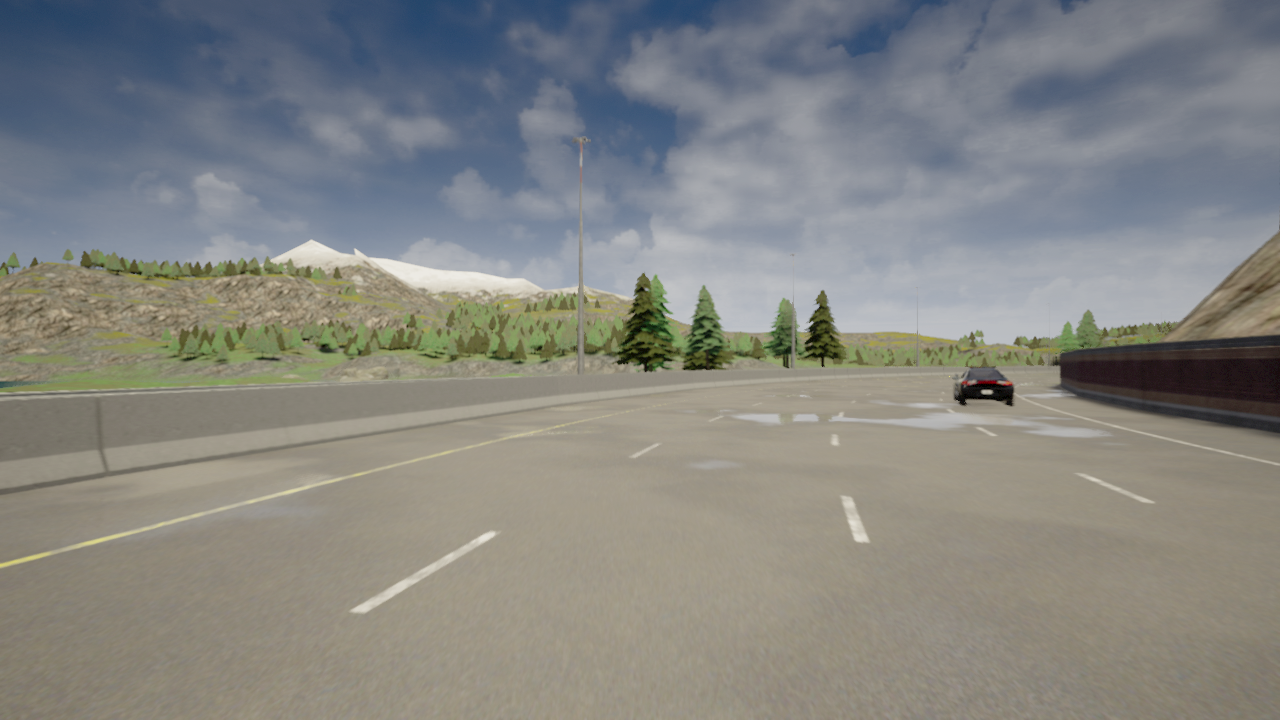}
        \caption*{ Transmitted Image}
    \end{minipage}
    \begin{minipage}[b]{0.48\linewidth}
        \centering
        \includegraphics[width=\linewidth]{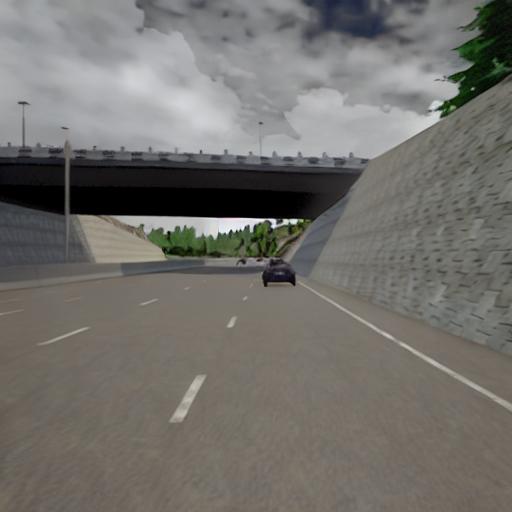}
        \caption*{ Reconstructed Image}
    \end{minipage}
    
    \caption{Examples of transmitted (left) and reconstructed (right) images at the receiver.}
    \label{fig:img_reconstruction}
\end{figure}

Figure~\ref{fig:img_reconstruction} presents example pairs of transmitted and reconstructed images. Although the regenerated images are not pixel-identical to the originals, they preserve the underlying scene semantics, including lane topology, road geometry, vehicle count, and inter-vehicle spatial relations. This confirms that the diffusion model does not perform appearance-level replication, but instead reconstructs the scene from a semantically grounded representation derived from the recovered graph. 

\subsection{Limitation}

Since the scene graph encodes only spatial relationships and high-level object properties, the reconstructed images are restricted to semantically grounded structural attributes. Contextual appearance attributes such as weather conditions, illumination, surface texture, or visual style are not preserved, as they are not embedded in the transmitted semantic representation.
In addition, although the decoder successfully regenerates spatial structure, it relies on CLIP-based textual conditioning for Stable Diffusion. The scene description grows proportionally with the number of detected objects and relations, and can exceed the maximum token length allowed by the CLIP text encoder. This forces truncation when the scene is dense, which may lead to missing semantic cues during reconstruction.


\section{Conclusion and Future Work}\label{Conclusion}
This paper presented an end-to-end semantic communication framework for vehicular environments in which scene understanding is transmitted in the form of structured semantics rather than raw visual data. By encoding images into graph-based representations that preserve object relations, the proposed system achieves a three-orders-of-magnitude reduction in payload size compared to conventional codecs while maintaining high semantic fidelity. At the receiver side, the recovered semantic graph enables both collision-risk estimation and image reconstruction through a diffusion-based semantic decoder, demonstrating dual task-level utility within a unified transmission pipeline. Experimental results showed that the framework remains robust under realistic channel impairments, with stable performance across varying SNR levels and a significantly lower FID score than state-of-the-art semantic communication methods.
\\
As part of future work, we plan to extend the semantic representation to incorporate environmental context such as weather, lighting, or road surface conditions, and to replace global text prompts with compact learned semantic embeddings to remove the dependency on CLIP token limits. Furthermore, another promising research direction is the joint training of the semantic encoder and diffusion-based decoder in an end-to-end manner, allowing task-oriented reconstruction to be optimized directly with respect to channel conditions.

\bibliographystyle{IEEEtran}

\bibliography{references.bib}

@article{zhang2022big,
  title={Big communications: Connect the unconnected},
  author={Zhang, Chuanting and Dang, Shuping and Alouini, Mohamed-Slim and Shihada, Basem},
  journal={Frontiers in Communications and Networks},
  volume={3},
  pages={785933},
  year={2022},
  publisher={Frontiers Media SA}
}

@article{kalor2024wireless,
  title={Wireless 6G connectivity for massive number of devices and critical services},
  author={Kalor, Anders E and Durisi, Giuseppe and Coleri, Sinem and Parkvall, Stefan and Yu, Wei and Mueller, Andreas and Popovski, Petar},
  journal={Proceedings of the IEEE},
  year={2024},
  publisher={IEEE}
}

@article{nguyen2022toward,
  title={Toward the age of intelligent vehicular networks for connected and autonomous vehicles in 6G},
  author={Nguyen, Van-Linh and Hwang, Ren-Hung and Lin, Po-Ching and Vyas, Abhishek and Nguyen, Van-Tao},
  journal={IEEE Network},
  volume={37},
  number={3},
  pages={44--51},
  year={2022},
  publisher={IEEE}
}

@inproceedings{ribouh2024semantic,
  title={Is semantic communication for autonomous driving secured against adversarial attacks?},
  author={Ribouh, Soheyb and Hadid, Abdenour},
  booktitle={2024 IEEE 6th International Conference on AI Circuits and Systems (AICAS)},
  pages={139--143},
  year={2024},
  organization={IEEE}
}

@article{getu2025semantic,
  title={Semantic communication: A survey on research landscape, challenges, and future directions},
  author={Getu, Tilahun M and Kaddoum, Georges and Bennis, Mehdi},
  journal={Proceedings of the IEEE},
  volume={112},
  number={11},
  pages={1649--1685},
  year={2025},
  publisher={IEEE}
}

@article{ribouh2025large,
  title={Large language model-based semantic communication system for image transmission},
  author={Ribouh, Soheyb and Saleem, Osama},
  journal={arXiv preprint arXiv:2501.12988},
  year={2025}
}

@article{lin2023embracing,
  title={Embracing AI in 5G-advanced toward 6G: A joint 3GPP and O-RAN perspective},
  author={Lin, Xingqin and Kundu, Lopamudra and Dick, Chris and Velayutham, Soma},
  journal={IEEE Communications Standards Magazine},
  volume={7},
  number={4},
  pages={76--83},
  year={2023},
  publisher={IEEE}
}

@article{lokumarambage2023wireless,
  title={Wireless end-to-end image transmission system using semantic communications},
  author={Lokumarambage, Maheshi U and Gowrisetty, Vishnu Sai Sankeerth and Rezaei, Hossein and Sivalingam, Thushan and Rajatheva, Nandana and Fernando, Anil},
  journal={IEEE Access},
  volume={11},
  pages={37149--37163},
  year={2023},
  publisher={IEEE}
}

@article{huang2022toward,
  title={Toward semantic communications: Deep learning-based image semantic coding},
  author={Huang, Danlan and Gao, Feifei and Tao, Xiaoming and Du, Qiyuan and Lu, Jianhua},
  journal={IEEE Journal on Selected Areas in Communications},
  volume={41},
  number={1},
  pages={55--71},
  year={2022},
  publisher={IEEE}
}

@inproceedings{tong2021federated,
  title={Federated learning based audio semantic communication over wireless networks},
  author={Tong, Haonan and Yang, Zhaohui and Wang, Sihua and Hu, Ye and Saad, Walid and Yin, Changchuan},
  booktitle={2021 IEEE Global Communications Conference (GLOBECOM)},
  pages={1--6},
  year={2021},
  organization={IEEE}
}

@article{liang2025semantic,
  title={Semantic communication for the internet of sounds: Architecture, design principles, and challenges},
  author={Liang, Chengsi and Sun, Yao and Thomas, Christo Kurisummoottil and Mohjazi, Lina and Saad, Walid},
  journal={IEEE Wireless Communications},
  year={2025},
  publisher={IEEE}
}

@article{sun2024task,
  title={Task-oriented scene graph-based semantic communications with adaptive channel coding},
  author={Sun, Shiqi and Qin, Zhijin and Xie, Huiqiang and Tao, Xiaoming},
  journal={IEEE Transactions on Wireless Communications},
  volume={23},
  number={11},
  pages={17070--17083},
  year={2024},
  publisher={IEEE}
}

@inproceedings{ribouh2020multiple,
  title={Multiple sequential constraint removal algorithm for channel estimation in vehicular environment},
  author={Ribouh, Soheyb and Elhillali, Yassin and Rivenq, Atika},
  booktitle={2020 International Symposium On Networks, Computers And Communications (ISNCC)},
  pages={1--7},
  year={2020},
  organization={IEEE}
}

@article{wang2025explicit,
  title={Explicit semantic-base-empowered communications for 6G mobile networks},
  author={Wang, Fengyu and Zheng, Yuan and Xu, Wenjun and Liang, Junxiao and Zhang, Ping and Han, Zhu},
  journal={Engineering},
  year={2025},
  publisher={Elsevier}
}

@article{zhou2023cognitive,
  title={Cognitive semantic communication systems driven by knowledge graph: Principle, implementation, and performance evaluation},
  author={Zhou, Fuhui and Li, Yihao and Xu, Ming and Yuan, Lu and Wu, Qihui and Hu, Rose Qingyang and Al-Dhahir, Naofal},
  journal={IEEE Transactions on Communications},
  volume={72},
  number={1},
  pages={193--208},
  year={2023},
  publisher={IEEE}
}

@article{zhang2024unified,
  title={A unified multi-task semantic communication system for multimodal data},
  author={Zhang, Guangyi and Hu, Qiyu and Qin, Zhijin and Cai, Yunlong and Yu, Guanding and Tao, Xiaoming},
  journal={IEEE Transactions on Communications},
  volume={72},
  number={7},
  pages={4101--4116},
  year={2024},
  publisher={IEEE}
}

@article{guo2025diffusion,
  title={Diffusion-driven semantic communication for generative models with bandwidth constraints},
  author={Guo, Lei and Chen, Wei and Sun, Yuxuan and Ai, Bo and Pappas, Nikolaos and Quek, Tony},
  journal={IEEE Transactions on Wireless Communications},
  year={2025},
  publisher={IEEE}
}

@article{liu2024ofdm,
  title={OFDM-based digital semantic communication with importance awareness},
  author={Liu, Chuanhong and Guo, Caili and Yang, Yang and Ni, Wanli and Quek, Tony QS},
  journal={IEEE Transactions on Communications},
  volume={72},
  number={10},
  pages={6301--6315},
  year={2024},
  publisher={IEEE}
}

@article{diao2025aligning,
  title={Aligning task-and reconstruction-oriented communications for edge intelligence},
  author={Diao, Yufeng and Zhang, Yichi and She, Changyang and Zhao, Philip Guodong and Li, Emma Liying},
  journal={IEEE Journal on Selected Areas in Communications},
  year={2025},
  publisher={IEEE}
}

@inproceedings{rombach2022high,
  title={High-resolution image synthesis with latent diffusion models},
  author={Rombach, Robin and Blattmann, Andreas and Lorenz, Dominik and Esser, Patrick and Ommer, Bj{\"o}rn},
  booktitle={Proceedings of the IEEE/CVF conference on computer vision and pattern recognition},
  pages={10684--10695},
  year={2022}
}

@inproceedings{cammerer2023neural,
  title={A neural receiver for 5G NR multi-user MIMO},
  author={Cammerer, Sebastian and A{\"\i}t Aoudia, Fay{\c{c}}al and Hoydis, Jakob and Oeldemann, Andreas and Roessler, Andreas and Mayer, Timo and Keller, Alexander},
  booktitle={2023 IEEE Globecom Workshops (GC Wkshps)},
  pages={329--334},
  year={2023},
  organization={IEEE}
}

@article{malawade2022roadscene2vec,
  title={roadscene2vec: A tool for extracting and embedding road scene-graphs},
  author={Malawade, Arnav Vaibhav and Yu, Shih-Yuan and Hsu, Brandon and Kaeley, Harsimrat and Karra, Anurag and Al Faruque, Mohammad Abdullah},
  journal={Knowledge-Based Systems},
  volume={242},
  pages={108245},
  year={2022},
  publisher={Elsevier}
}

@data{c0z9-1p30-21,
doi = {10.21227/c0z9-1p30},
url = {https://dx.doi.org/10.21227/c0z9-1p30},
author = {Brandon Hsu and Shih-Yuan Yu and Arnav Malawade and Deepan Muthirayan and Pramod Prabhakar Khargonekar and Mohammad Abdullah Al Faruque},
publisher = {IEEE Dataport},
title = {SceneGraph-Risk-Assessment dataset},
year = {2021} }

@article{dosovitskiy2017carla,
  title={" CARLA: An open urban driving simulator", Conference on Robot Learning, PMLR},
  author={Dosovitskiy, A and Ros, G and Codevilla, F and Lopez, A and Koltun, V},
  year={2017}
}

@article{hoydis2022sionna,
  title={Sionna: An open-source library for next-generation physical layer research},
  author={Hoydis, Jakob and Cammerer, Sebastian and Aoudia, Fay{\c{c}}al Ait and Vem, Avinash and Binder, Nikolaus and Marcus, Guillermo and Keller, Alexander},
  journal={arXiv preprint arXiv:2203.11854},
  year={2022}
}

@article{getu2023making,
  title={Making sense of meaning: A survey on metrics for semantic and goal-oriented communication},
  author={Getu, Tilahun M and Kaddoum, Georges and Bennis, Mehdi},
  journal={IEEE Access},
  volume={11},
  pages={45456--45492},
  year={2023},
  publisher={IEEE}
}

@article{wu2025pragmatic,
  title={A Pragmatic Note on Evaluating Generative Models with Fr$\backslash$'echet Inception Distance for Retinal Image Synthesis},
  author={Wu, Yuli and Liu, Fucheng and Yilmaz, R{\"u}veyda and Konermann, Henning and Walter, Peter and Stegmaier, Johannes},
  journal={arXiv preprint arXiv:2502.17160},
  year={2025}
}

@inproceedings{yang2024sg2sc,
  title={SG2SC: A generative semantic communication framework for scene understanding-oriented image transmission},
  author={Yang, Minxi and Gao, Dahua and Xie, Feng and Li, Jiaxuan and Song, Xiaodan and Shi, Guangming},
  booktitle={ICASSP 2024-2024 IEEE International Conference on Acoustics, Speech and Signal Processing (ICASSP)},
  pages={13486--13490},
  year={2024},
  organization={IEEE}
}

@article{grassucci2023generative,
  title={Generative semantic communication: Diffusion models beyond bit recovery},
  author={Grassucci, Eleonora and Barbarossa, Sergio and Comminiello, Danilo},
  journal={arXiv preprint arXiv:2306.04321},
  year={2023}
}

@inproceedings{yang2023witt,
  title={WITT: A wireless image transmission transformer for semantic communications},
  author={Yang, Ke and Wang, Sixian and Dai, Jincheng and Tan, Kailin and Niu, Kai and Zhang, Ping},
  booktitle={ICASSP 2023-2023 IEEE International Conference on Acoustics, Speech and Signal Processing (ICASSP)},
  pages={1--5},
  year={2023},
  organization={IEEE}
}

\end{document}